# Detection of Geometric Phases in Superconducting Nanocircuits

Giuseppe Falci[(1,4)], Rosario Fazio[(1,4)],G. Massimo Palma[(2,4)], Jens Siewert[(1,4)], and VlatkoVedral[(3)]

[(1)]*Dipartimento di Metodologie Fisiche e Chimiche (DMFCI), Università di Catania, viale A.Doria 6, I-95125 Catania, Italy*

[(2)]*Dipartimento di Scienze Fisiche ed Astronomiche, (DSFA),Università di Palermo, via Archirafi 36, I-90123 Palermo, Italy*

[(3)]*Centre for Quantum Computation, Clarendon Laboratory, University of Oxford, Parks Road, Oxford OX1 3PU, UK*

[(4)]*Istituto Nazionale per la Fisica della Materia (INFM), Unità di Catania e Palermo.*

**When a quantum mechanical system undergoes an adiabatic cyclic evolution it acquires a geometrical phase factor [1] in addition to the dynamical one. This effect has been demonstrated in a variety of microscopic systems [2]. Advances in nanotechnologies should enable the laws of quantum dynamics to be tested at the macroscopic level [3], by providing controllable artificial two-level systems ( for example, in quantum dots [4] and superconducting devices [5,6]). Here we propose an experimental method to detect geometric phases in a superconducting device. The setup is a Josephson junction nanocircuit consisting of a superconducting electron box. We discuss how interferometry based on geometrical phases may be realized, and show how the effect may applied to the design of gates for quantum computation.**

Interferometry based on Berry's phase has been proposed to realize quantum gates. This is at the heart of geometric quantum computation that has been implemented in NMR [7]. Quantum computation based on non-Abelian phases has been proposed [8]. Quantum computers can be viewed as programmable quantum interferometers [9]. Initially prepared in a superposition of all the possible input states, the computation evolves in parallel along all its possible paths, which will interfere constructively towards the desired output state. This intrinsic parallelism in the evolution of a quantum system allows for exponentially more efficient ways of performing computation [10].

The quest for large scale integrability has stimulated an increasing interest in superconducting nanocircuits [11-16] as possible candidates for the implementation of a quantum computer. Mesoscopic Josephson junctions can be prepared in a controlled superposition of charge states [17,18] and the coherent time evolution in a Josephson charge qubit has been recently observed [5]. These are the first important experimental steps towards the implementation of a solid state quantum computer.

An important aspect of this research activity is that it unveils fundamental problems related to the quantum mechanical behavior of macroscopic systems. Here we study quantum interferometry based on geometric phases in mesoscopic Josephson devices. A well known example in classical superconductivity is the Aharonov-Bohm (AB) effect which manifests itself, for example, in the periodic modulation of the critical current in classical SQUIDs (ref.19). The AB effect provides evidence of the macroscopic coherence of the superconducting condensate. We will show that it is possible to detect geometric phases, other than the AB phase, in the coherent dynamics of superconducting nanocircuits. Quantum interferometry based on this geometric phases allows to develop a new design of gates for quantum computation using charge qubits (the designs proposed up to now for superconducting quantum gates use interferometry of dynamic phases). Indeed, it is possible to introduce conditional geometric phases, that is, in which the geometric phase on one qubit is controlled by the state of the neighboring one. Such a feature, in itself a remarkable effect in the theory of geometric phases, is an important requirement of quantum computation.

The setup we consider is shown in Fig.1a. It consists of a superconducting electron box formed by an asymmetric SQUID, pierced by a magnetic flux $\Phi$ and with an applied gate voltage $V_x$. The device operates in the charging regime, that is, the Josephson couplings $E_{J1(2)}$ of the junctions are much smaller than the charging energy $E_{ch}$. We further assume that the temperature is much lower than $E_{J1(2)}$. The hamiltonian is [19]

$$H = E_{ch}(n-n_x)^2 - E_J(\Phi) \cos(\theta - \alpha) \qquad (1)$$

where

$$E_J(\Phi) = \sqrt{(E_{J1} - E_{J2})^2 + 4E_{J1}E_{J2}\cos^2\left(\pi\frac{\Phi}{\Phi_0}\right)} \qquad \tan\alpha = \frac{E_{J1} - E_{J2}}{E_{J1} + E_{J2}}\tan\left(\pi\frac{\Phi}{\Phi_0}\right)$$

and $\Phi_0 = h/2e$ is the (superconducting) quantum of flux. The phase difference across the junction $\theta$ and the number of Cooper pairs n are canonically conjugate variables $[\theta, n] = i$. The parameters of the hamiltonian can be controlled. The offset charge $2en_x$ can be tuned by changing $V_x$ (see Fig.1a) and the coupling $E_J(\Phi)$ depends on $\Phi$, as in the device proposed in Ref.[11,12]. We propose the use of an asymmetric SQUID, which permits to control the phase shift $\alpha(\Phi)$ as well.

At temperatures much lower than $E_{ch}$, if $n_x$ varies around the value 1/2, only the two charge eigenstates n=1,0 are important. They constitute the basis $\{|0>, |1>\}$ of the computational Hilbert space of the qubit [11-13]. The effective hamiltonian is obtained by projecting Eq.(1) on the computational Hilbert space, and reads $H_B = -(1/2)\boldsymbol{B}\cdot\boldsymbol{\sigma}$,

where we have defined the fictitious field $\mathbf{B} \equiv (E_J\cos\alpha, -E_J\sin\alpha, E_{ch}(1-2n_x))$ and $\sigma$ are the Pauli matrices.

The system thus behaves an artificial spin in a magnetic field. Charging couples the system to $B_z$ whereas the Josephson term determines the projection in the $xy$ plane. By changing $V_x$ and $\Phi$ the qubit hamiltonian $H_B$ describes a cylindroid in the parameter space $\{\mathbf{B}\}$. The presence of higher charge states leads to quantum leakage. This effect can be minimized by a fine tuning of the parameters of the devices [16].

We now consider the adiabatic time evolution of the qubit hamiltonian. By changing $H_B$ around a circuit in the parameter space $\{\mathbf{B}\}$, the eigenstates will accumulate a Berry phase proportional to the solid angle that the circuit subtends at the degeneracy point, $\mathbf{B}=0$. The Berry phase $\gamma_B(\Phi_M, n_{xm})$ accumulated by the two lowest eigenstates along a circuit where $n_x$ is varied from $n_{xm}$ to $1/2$ and the flux from zero to $\Phi_M$ (see the inset of Fig.1a) is plotted in Fig.2.

Notice that non-trivial loops with a controllable Berry phase are possible in our setup thanks to the asymmetry in the SQUID. In the symmetric case $\alpha(\Phi)=0$ and the evolution takes place in a strip of the plane $B_y=0$; that is, the Berry phase is zero. The distinct feature of the (geometric) interferometry proposed here is that the geometric phase does depend on *both* gate voltage and flux. It differs from "classical" interference in mesoscopic junctions, achieved either by changing the magnetic field (in the flux regime) or by a gate voltage modulation (in the charge regime).

The experimental setup required to measure $\gamma_B(\Phi_M, n_{xm})$ is already available and corresponds to that of ref.5. In the following, we describe a procedure to measure the Berry phase. The system is prepared in the ground state of the Hamiltonian at $n_x=0$ and $\Phi=0$ and then a sudden switching to the point $n_x=1/2$ and $\Phi=0$ ($\mathbf{B}_0 = (E_J(0),0,0)$) is applied to produce the initial hamiltonian $H_{B0}$. The initial charge state is then a linear superposition of the eigenstates $\{|B_{0+}\rangle, B_{0-}\rangle\}$ of the hamiltonian $H_{B0}$. This is analogous to the splitting of the photon wavefuntion at the first beamsplitter of a Mach-Zender interferometer. A phase difference between the states $|B_{0+/-}\rangle$ can be introduced by adiabatically dragging $H_B$ along a closed loop (see the inset of Fig1a). The phases acquired this way will have both a geometrical and a dynamical component. In order to eliminate the latter it is sufficient to perform a NOT gate, for instance by applying an ac gate voltage pulse that swaps the eigenstates. If the same loop as before is covered backwards by $H_B$, the geometrical component of the phase will add while the dynamical phases will cancel each other [7]. The final step is to measure the charge state of the qubit. The probability to measure a charge 2e (n=1) in the box at the end of this procedure is given by (up to negligible terms of the order of $[E_J(\Phi_0/2)/E_{ch}]^2$)

$$P(1) = \sin^2(2\gamma_B) \qquad (2)$$

*independently* on the time elapsed. The result given in Eq.(2) holds as long as the second adiabatic loop retraces exactly the first one, in order to cancel the dynamical phase. If this is not the case, the measurement will be affected by random errors that will reduce the amplitude of the interference fringes. The dephasing time should be longer than the time needed for the adiabatic manipulation, which has to be longer than the typical time scale for the dynamics, $h/|E_{J1} + E_{J2}|$. The results of ref.5 indicate that with the present technology there is already sufficient control of the dynamics to detect the geometric phase.

Geometric interferometry in mesoscopic superconducting devices can be conveniently discussed in the langauge of quantum computation. Conditional geometric interferometry, which corresponds to the implementation of a universal two-qubit gate is realized by coupling capacitively two asymmetric SQUIDS (see Fig.1b). All these ingredients provide a framework for the implementation of geometric quantum computation in a solid state device. If the coupling capacitance $C_K$ is smaller than the others the hamiltonian is

$$H_G = \sum_{i=1,2} H_i + \delta E (n_1 - n_{x,1})(n_2 - n_{x,2}) \qquad (3)$$

where $H_i$ refer to the two uncoupled qubits defined in Eq.(1) and $\delta E = 2E_{ch}C_K/C$ is the charging energy which derives from the capacitive coupling. Gate voltages and magnetic fluxes can be independently fixed for the two qubits. This setup can be used to implement a controlled phase shift; that is, the geometric phase of one qubit (the target) depends on the state of the neighboring one (the control qubit) and to detect it. Because of universality in quantum computation [21], this gate can be used to perform any computational task, if complemented with the single qubit operations described above. The gate voltage of the control qubit is kept away from the resonance condition. As a result of the coupling, the effective charging of the target qubit, and consequently the surface spanned in the parameter space, depends on the state of the control qubit. This implies that the Berry phase $\gamma_B$ is conditional. In the basis of the charge states the unitary operator that describes this gate is

$$\begin{pmatrix} \exp(-i\gamma_0) & 0 & 0 & 0 \\ 0 & \exp(i\gamma_0) & 0 & 0 \\ 0 & 0 & \exp(-i\gamma_1) & 0 \\ 0 & 0 & 0 & \exp(i\gamma_1) \end{pmatrix}$$

where $\gamma_i$ is the Berry phase when the state of the control qubit is i=0,1. Measurement of the conditional phase shift can be performed following the procedure outlined previously. One important issue is that the NOT operation needed to cancel the dynamical phase can be performed with an ac voltage bias pulse whose amplitude does not depend on the value of the control qubit. As it is not possible to switch off the

Josephson coupling of the control qubit completely, the real gate will have very small off-diagonal terms that are irrelevant on the time scale of manipulation.

Quantum interferometry provides a clear conceptual framework to describe in a unified language a broad range of quantum phenomena and to transfer ideas for new experimental investigation of fundamental aspects of quantum theory. We have described a new class of quantum geometric phase effects in small Josephson junctions, other than conventional AB effect. Adiabatic pumping of Cooper pairs was interpreted in terms of Berry phase in ref.22. We discuss how to realize interferometry based on geometrical phases and show how this effect can find application in the design of new gates for quantum computation. The unifying concept underlying these two apparently distinct topics is the fact that both the detection procedure of geometric phases and quantum computation can be described in interferometric terms.

**Acknowledgements.** The authors would like to thank D.V.Averin, A.Ekert, G. Giaquinta, J. Jones, B.Pannetier and E.Paladino for helpful discussions. G.F. acknowledges kind hospitality at LEPES-CNRS (Grenoble). Part of the work of R.F. was done at ISI-Torino. This work was supported by the European Community (TMR, IST-SQUIBIT, IST-EQUIP), by INFM-PRA-SSQI and by ELSAG S.p.A.

**Correspondence and requests for materials should be addressed to G. Falci (e-mail: gfalci@dmfci.ing.unict.it).**

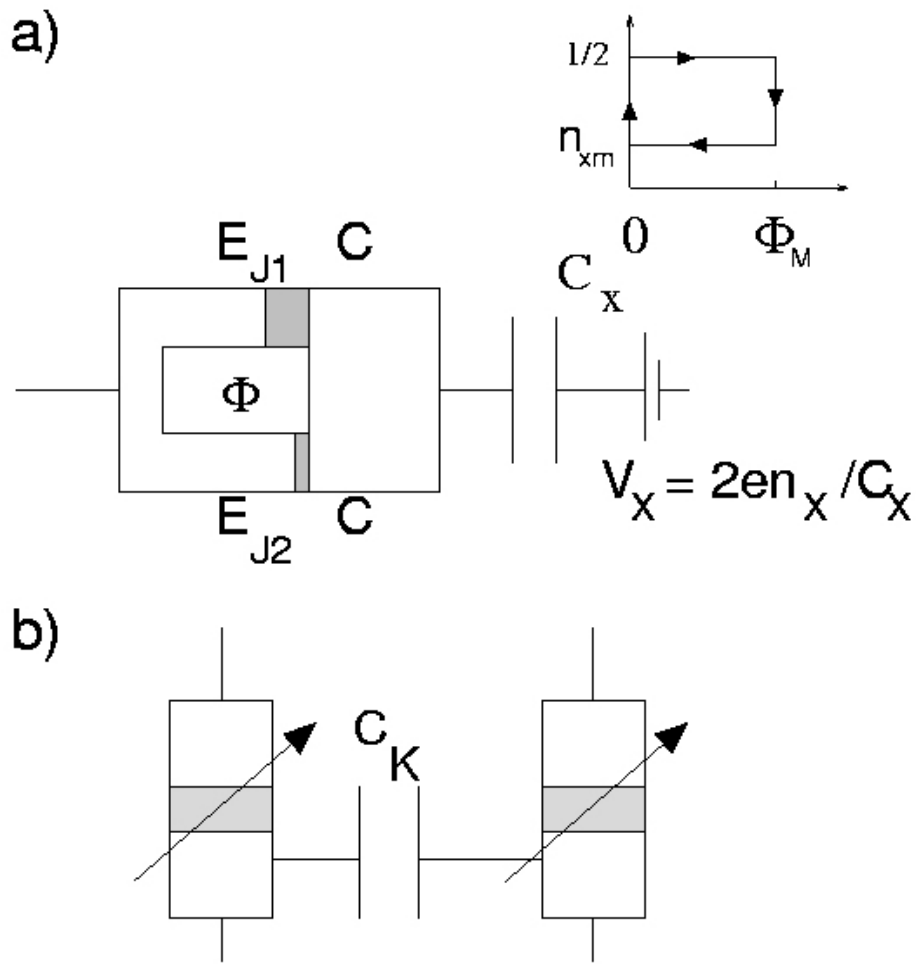

Figure 1. Schematic design of the system and of the gates. a) The system consists of a superconducting electron box formed by an asymmetric SQUID, pierced by a magnetic flux $\Phi$, and with an applied gate voltage $V_x$. The offset charge $n_x$ defined in the paper is proportional to $V_x$. The device operates in the charge regime, that is, $E_{J1}, E_{J2} \ll E_{ch}$. The cyclic evolution of the hamiltonian is obtained by changing the gate voltage and the magnetic flux along the path shown in the inset. b) Schematic design of the two-bit gate. The two qubits are coupled capacitively.

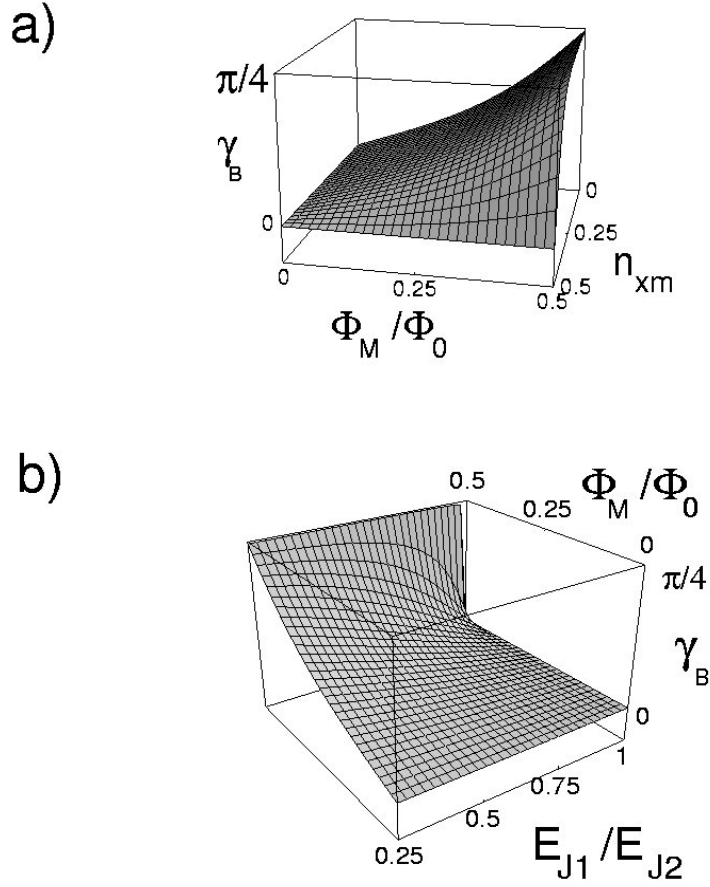

Figure 2. Geometric phase produced by the adiabatic manipulations. a)The Berry phase is calculated for the loop shown in the inset of Fig1a as a function of the flux and the external charge ( $\gamma_B(\Phi_M, n_{xm}) = \pm [E_{ch}(1-2n_{xm})/4(E_{J1}E_{J2})^{1/2}]$ $\lambda\mu\ \Pi(\pi\Phi_M/\Phi_0, 1-\mu^2, \lambda)$ where $\Pi(x,n,k)$ is the elliptic function of the third kind [20], $\lambda^2 = 4E_{J1}E_{J2}/[(E_{ch}(1-2n_{xm}))^2 + (E_{J1} + E_{J2})^2]$ and $\mu = (E_{J1} - E_{J2})/(E_{J1} + E_{J2})$ ). In this plots the parameters are $E_{J1}= 0.25\ E_{J2}$ and $E_{ch}=5\ (E_{J1} + E_{J2})$. b) The Berry phase is plotted as a function of the asymmetry parameter $E_{J1}/E_{J2}$ and the flux for the path where $n_{xm}=0$.